\documentclass[prl,twocolumn,amsmath,amssymb,floats,showpacs]{revtex4-1}
\usepackage{graphicx}
\usepackage[caption=false]{subfig}
\usepackage{amsmath}
\usepackage{amsfonts}
\usepackage{epstopdf}
\usepackage{amsbsy}
\usepackage{color}
\usepackage{dcolumn}
\usepackage{bm}

\newcommand{\bk}{\boldsymbol k}
\newcommand{\br}{\boldsymbol r}
\newcommand{\Spin}{\boldsymbol S}
\newcommand{\pauli}{\boldsymbol \tau}
\newcommand{\exch}{\mathcal J}
\newcommand{\s}{\sigma}

\begin{document}

\title{Tunable magnetic anisotropy from higher-harmonics exchange scattering on the surface of a topological insulator}
\author{Jens Paaske and Erikas Gaidamauskas}
\affiliation{Center for Quantum Devices, Niels Bohr Institute,
University of Copenhagen, Universitetsparken 5, DK-2100 Copenhagen,
Denmark}

\date{\today}

\begin{abstract}
We show that higher-harmonics exchange scattering from a magnetic adatom on the surface of a three dimensional topological insulator leads to a magnetic anisotropy whose magnitude and sign may be tuned by adjusting the chemical potential of the helical surface band. As chemical potential moves from the Dirac point towards the surface band edge, the surface normal is found to change from magnetic easy, to hard axis. Hexagonal warping is shown to diminish the region with easy axis anisotropy, and to suppress the anisotropy altogether. This indirect contribution can be comparable in magnitude to the intrinsic term arising from crystal field splitting and atomic spin-orbit coupling, and its tunability with chemical potential makes the two contributions experimentally discernible, and endows this source of anisotropy with potentially interesting magnetic functionality.
\end{abstract}


\maketitle

\begin{figure}[t]
\includegraphics[width=\columnwidth]{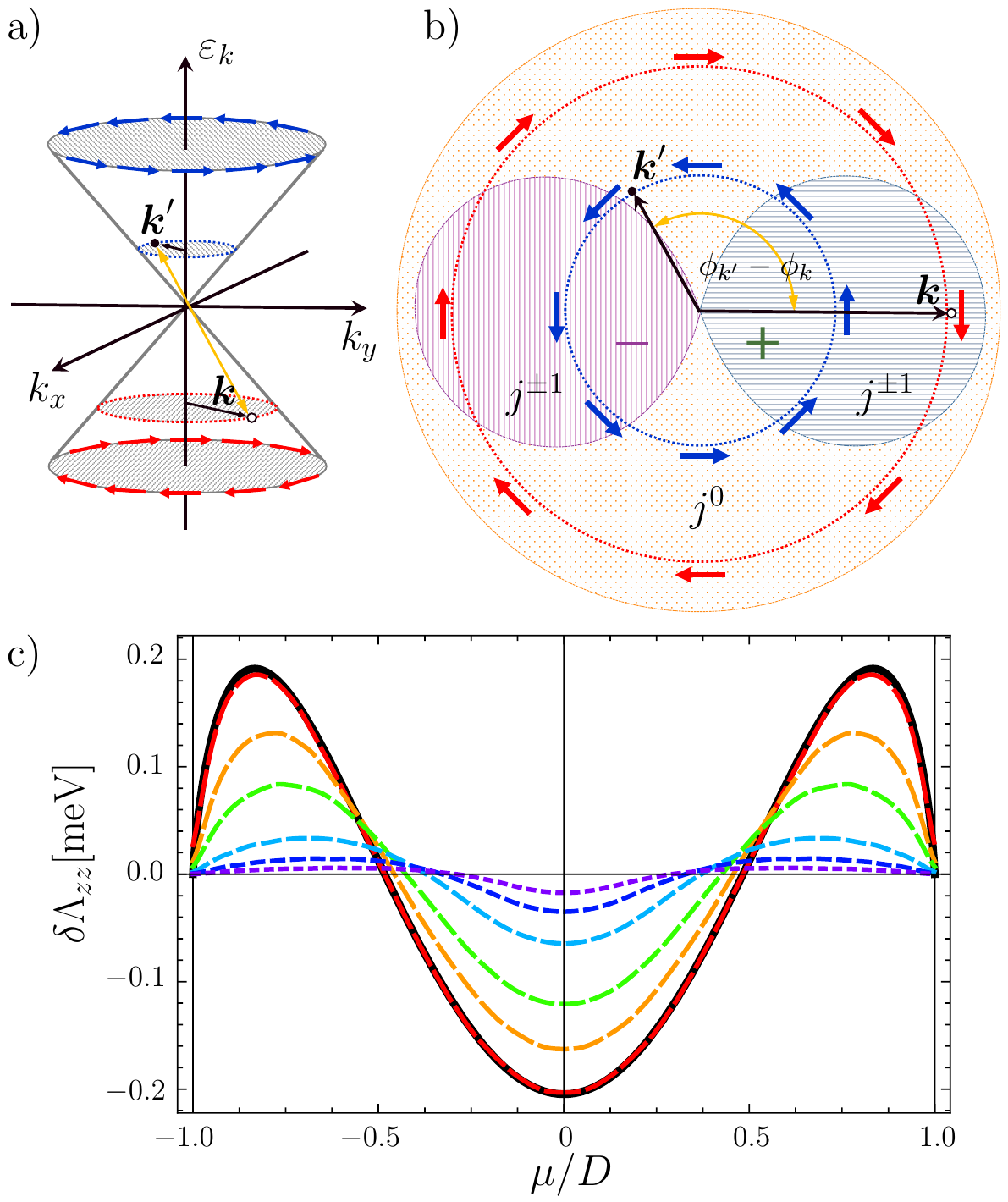}
\caption{(Color online) a) The conical TSS band with an indication of the opposite helicity of positive- and negative-energy states, and an interband scattering process from $\bk$ to $\bk'$. b) Illustration of $s$-wave (orange/dotted) and $p$-wave (purple-green/striped) scattering amplitudes in Eq.~\eqref{eq:Lamb} and their relation to the scattering angle for the interband scattering process shown in panel a. c) Anisotropy energy resulting from two constant amplitudes $2j^{1}=j^{0}= 10$ eV \AA$^{2}$ and using estimated values for pristine Bi$_2$Se$_3$ surface states with $v_{F}\approx 3000$ meV \AA, and half bandwidth $D\approx400$ meV (where the built-in chemical potential lies around 300 meV)~\cite{Scholz2012, Schlenk2013}. The thick black curve shows the perturbative result in Eq.~\eqref{eq:res1}. Curves in red to magenta (and smaller dashing) correspond to stronger warping, with $\lambda=\{0, 1, 2, 5, 10, 20\}\times10^5$ meV$^2$\AA$^3$, calculated by numerical $k$-space diagonalization.}
\label{fig:cone}
\end{figure}

A ubiquitous signature of three-dimensional topological insulators (TI) is their helical surface states~\cite{Hasan2010, Qi2011}. On a given surface, these are topologically protected electronic states with a linear dispersion relation in the form of a single Dirac cone centered at the $\Gamma$-point. Due to the underlying spin-orbit coupling, these topological surface states (TSS) have spin locked to momentum, and  positive, and negative energy states, come with opposite helicity, i.e. opposite winding directions of the in-plane electron spin around the Dirac cone~\cite{Hsieh2009, Wang2011}(cf. Fig.~\ref{fig:cone}a). A spin polarisation perpendicular to the surface will mix the two helicities, open a gap in the Dirac spectrum~\cite{Wray2011, Xu2012, Lee2015}, and thereby give rise to a number of interesting phenomena like anomalous quantum Hall effect, quantized Faraday and Kerr effects, magneto-electric effects, and charge-induced magnetic monopoles~\cite{Hasan2010, Tse2010, Liu2008, Liu2015}.

The macroscopic spin polarization, required for this gap opening, can be induced either by proximity to a ferromagnetic insulator or more directly by doping with magnetic atoms~\cite{Wray2011, Xu2012, Lee2015}. However, since the polarization needs to be perpendicular to the surface, it is important to understand and control the nature of the single-ion magnetic anisotropy, which is to be expected for magnetic moments located near the surface. This question has been addressed experimentally by means of x-ray magnetic circular dichroism (XMCD), and initial measurements found the Bi$_2$Se$_3$ surface to constitute a magnetic easy plane for Fe adatoms, with an anisotropy barrier of $K=1.9$ meV per atom~\cite{Honolka2012}. In contrast, a magnetic easy axis normal to the surface was later observed for Fe on Bi$_2$Se$_3$~\cite{Ye2013}, and more recently for Fe on Bi$_2$Te$_3$ with a barrier as high as $K=10$ meV per atom~\cite{Eelbo2014}. As summarized in Ref.~\onlinecite{Eelbo2014}, sample preparation appears to be decisive, both for the ratio of spin, to orbital magnetic moment, and for the magnetic anisotropy.

Density functional theory (DFT) has provided a consistent explanation for the observations in Ref.~\onlinecite{Eelbo2014}, whereas dynamical hybridization effects had to be invoked to explain the easy-plane anisotropy found in Ref.~\onlinecite{Honolka2012}. The {\it intrinsic} magnetic anisotropy determined this way relies on crystal field splitting combined with spin-orbit coupling on the impurity atom, thus taking into account the microscopic details of the specific adatom accommodation on the surface. However, a TI surface provides for a second source of {\it indirect} magnetic anisotropy which relies on the helical nature of the TSS themselves and is therefore not captured by a DFT calculation unless it includes enough of the bulk TI to provide for the TSS. This was first pointed out by N\~unez et al.~\cite{Nunez2012} who showed that a finite-range exchange coupling between a magnetic impurity and a half-filled Dirac band leads to a substantial single-ion magnetic anisotropy with an easy axis normal to the surface, consistent with the findings of Refs.~\onlinecite{Ye2013, Eelbo2014}.

Here we show that such higher-harmonics exchange scattering gives rise to a magnetic anisotropy which depends strongly on the chemical potential of the helical TSS and even changes sign some distance away from half-filling, indicating a transition from easy-axis, to easy-plane anisotropy. This adds an important ingredient to the resolution of the experimental situation, not least since the chemical potential for the TSS is found to move as a result of the adatom deposition~\cite{Scholz2012, Honolka2012, Schlenk2013}. This provides a new indirect link between concentration (and possible thermally activated subsurface diffusion) of adatoms, and their resulting magnetic anisotropy.


{\it Model and theory.} We consider a single magnetic impurity placed on the surface of a 3D-topological insulator. The two-dimensional helical TSS are modelled by the Hamiltonian

\begin{align}
H_{el} = \sum_{\bk, \sigma'\sigma}
c^{\dagger}_{\bk \sigma'}{\boldsymbol h}(\bk)\cdot\pauli_{\sigma'\sigma}
c^{}_{\bk \sigma},\label{eq1}
\end{align}
where $\pauli$ denotes the vector of Pauli-matrices and
${\boldsymbol h}(\bk)=\left(-v_{F}k\sin\phi_{\bk}, v_{F}k\cos\phi_{\bk},\lambda k^{3}\cos3\phi_{\bk}\right)$, where $\phi_{\bk}$ denotes the polar angle of the momentum vector $\bk$ and $\lambda$ sets the strength of the hexagonal warping~\cite{Lee2009,Fu2009}. The eigenvalues of $\mathcal{H}$ are indexed by {\it helicity} $\eta=\pm 1$ and given by $\varepsilon_{\eta\bk}=\eta|{\boldsymbol h}|=\eta\sqrt{v_{F}^{2}k^{2}+\lambda^{2}k^{6}\cos^{2}(3\phi_{\bk})}$, which is isotropic for $k\ll k_{w}$ where $k_{w}=\sqrt{v_{F}/\lambda}$.
Within this continuum model, we assume the impurity spin to be exchange coupled to the TSS by the $sd$-Hamiltonian,
\begin{align}
\label{eq2}
H_{sd}
=\sum_{\s\s'}\int d^{2}r \exch(r)\Spin\cdot\pauli_{\s\s'}
\psi^{\dagger}_{\s} (\br)\psi_{\s'}(\br),
\end{align}
in terms of a finite-range isotropic exchange function $\exch(r)$. In momentum space this Hamiltonian reads $H_{sd}
=\sum_{\bk' \bk,\sigma' \sigma}{\tilde\exch}_{\bk'-\bk} \boldsymbol S \cdot \boldsymbol \tau_{\sigma' \sigma}  c^{\dagger}_{\bk' \sigma'} c_{\bk \sigma}$, with
\begin{align}
{\tilde\exch}_{\bk'-\bk}=\int_{0}^{\infty}\!\!dr\,r \exch(r)J_{0}(r|\bk'-\bk|),
\end{align}
given in terms of the zeroth Bessel functions, $J_{0}$. The two momenta may be separated by the following expansion of the zeroth Bessel function~\cite{Lebedev1972}:
\begin{align}
{\tilde\exch}_{\bk'-\bk}=\sum_{l=-\infty}^{\infty}j^{l}_{k'k}e^{il(\phi'-\phi)},
\end{align}
where $\phi (\phi')$ denotes the polar angle of the incoming (outgoing) momentum, and the $l$'th harmonic has weight $j^{l}_{k'k}=\int_{0}^{\infty}\!dr\,r \exch(r)J_{l}(r k')J_{l}(r k)$.
Integrating out the TSS electrons to second order in the exchange coupling, akin to deriving the Ruderman-Kittel-Kasuya-Yosida (RKKY) interaction~\cite{Yosida1996,Zhu2011}, now leads to the following effective Hamiltonian for the impurity spin,
\begin{align}\label{eq:Heff}
H_{\rm eff}=\!\!\sum_{i,j=x,y,z}\!\!\Lambda_{ij} S^i S^j,
\end{align}
with the tensor $\Lambda_{ij}$ given as
\begin{align}\label{eq:Lamb}
\Lambda_{ij}=&\,
\frac{1}{2}
\int\!\frac{d^{2}k'}{(2\pi)^2}
\int\!\frac{d^{2}k}{(2\pi)^2}
\chi_{ij}(\bk',\bk) \nonumber\\
&\hspace*{15mm}\times\sum_{l_1 l_2} j^{l_1}_{k'k}j^{l_2}_{k k'}e^{i(l_{1}-l_{2})(\phi'-\phi)},
\end{align}
in terms of the electronic spin-susceptibility tensor
\begin{align}\label{eq:suscept}
\chi_{ij}(\bk',\bk)=&\,
\int_{0}^{\beta}\!\!\!d\tau\,
{\rm Tr}[\tau^i \mathcal{G}^{(0)}(\bk', \tau)\tau^j \mathcal{G}^{(0)}(\bk, -\tau)]
\end{align}
with non-interacting TSS Matsubara Green functions\begin{eqnarray}
\lefteqn{\mathcal{G}^{(0)}_{\sigma \sigma'}(\bk, \tau)=
\sum_{\eta=\pm}\frac{1}{2}
\left[\tau^{0}+\eta\pauli\cdot{\hat{\boldsymbol h}}(\bk)\right]_{\sigma \sigma'}}\\
&\times
\left[
n_F(\xi_{\eta\bk})\Theta(-\tau)+
(n_F(\xi_{\eta\bk})-1)\Theta(\tau)\right]e^{-\xi_{\eta\bk}\tau},\nonumber
\end{eqnarray}
where $n_{F}$ denotes the Fermi function.
A chemical potential for the TSS electrons has been included through $\xi_{\eta\bk}=\varepsilon_{\eta\bk}-\mu$, and ${\hat{\boldsymbol h}}$ denotes the unit vector along ${\boldsymbol h}$. Altogether, this yields the diagonal tensor
\begin{align}
{\underline{\underline\Lambda}}=
\left(
  \begin{array}{ccc}
    \Lambda_{0} & 0 & 0 \\
    0 & \Lambda_{0} & 0 \\
    0 & 0 & \Lambda_{0}+\delta\Lambda_{zz} \\
  \end{array}
\right),
\end{align}
with an irrelevant isotropic term $\Lambda_{0}$ and a longitudinal anisotropy $\delta\Lambda_{zz}$ given by
\begin{eqnarray}\label{eq:dLamb}
\lefteqn{\delta\Lambda_{zz}=\,
-\frac{1}{8\pi^{2}}
\int\!dk' k'
\int\!dk k
\int_{0}^{\pi}\!\frac{d\phi}{\pi}\,
\int_{0}^{\pi}\!\frac{d\phi'}{\pi}\zeta(\bk,\bk')}\\
&\times\frac{v_{F}^{2}k k'}{\varepsilon_{+\bk}\varepsilon_{+\bk'}}
\left(
\Gamma^{1}_{kk'}-\left(\frac{2k k'}{k_{w}^{2}}\right)^{2}
\Gamma^{3}_{kk'}\cos\phi\cos\phi'\cos(\phi'-\phi)
\right)\nonumber,
\end{eqnarray}
where $\Gamma^{n}_{kk'}=\sum_{l=-\infty}^{\infty}j^{l}_{k'k}j^{l+n}_{k'k}$, and with $\mu$ entering via
\begin{align}
\zeta(\bk,\bk')
=\!\!\sum_{\eta,\eta'=\pm 1}\!\!\eta\eta'\frac{n_{F}(\varepsilon_{\eta\bk}-\mu)-n_{F}
(\varepsilon_{\eta'\bk'}-\mu)}{\varepsilon_{\eta\bk}-\varepsilon_{\eta'\bk'}}.
\end{align}


{\it Numerical results and discussion.} The basic mechanism for the anisotropy is the non-trivial dependence of ${\tilde\exch}_{\bk'-\bk}$ on the scattering angle, $\phi'-\phi$, which introduces a relative sign between spin-flip, and non-flip scattering, as illustrated for an interband ($|\bk,{\textcolor{red} \downarrow}\rangle\to|\bk',{\textcolor{blue} \sigma}\rangle$) transition in Fig.~\ref{fig:cone}b. Technically speaking, the momentum dependence from the higher harmonics, $e^{il(\phi'-\phi)}$, leads to a non-zero contribution from the term proportional to $\pauli\cdot{\hat{\boldsymbol h}}(\bk)$ in $\mathcal{G}^{(0)}$, which in turn responds differently to the subsequent trace with respectively $\tau^{x,y}$ and $\tau^{z}$ in Eq.~\eqref{eq:suscept}. At $\mu=T=0$, only interband processes are possible, and one finds that $\zeta(\bk,\bk')=2/(\varepsilon_{+\bk}+\varepsilon_{+\bk'})$, implying that $\delta\Lambda_{zz}<0$, i.e. a preferred orientation for the impurity spin colinear with the $z$-direction.

\begin{figure}[t]
\includegraphics[width=\columnwidth]{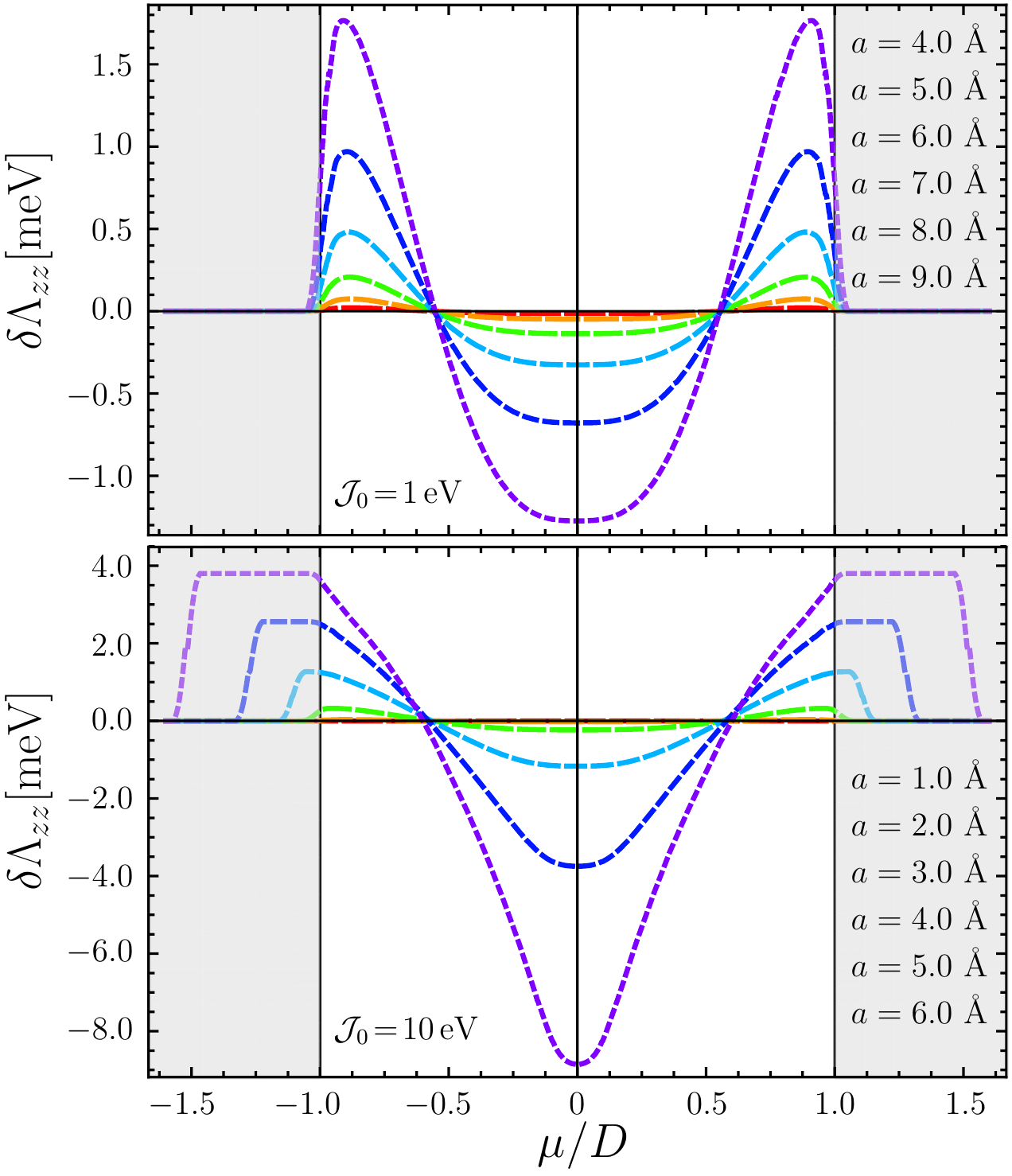}
\caption{(Color online) Plots of the longitudinal magnetic anisotropy energy for an isotropic step-function impurity potential, $\exch_{0}(r)$, with varying ranges $a$ and fixed amplitudes $\exch_{0}=1$ eV (upper panel) and $\exch_{0}=10$ eV (lower panel). Curves in red to magenta (and smaller dashing) correspond to larger values of $a$, as listed in the insets.
}
\label{fig:mucombi}
\end{figure}

As a first approximation, we simply neglect the momentum dependence of the couplings $j^{l}$ and leave out the effects of warping by setting $\lambda=0$. At zero temperature, the remaining integral yields
\begin{align}
\delta\Lambda_{zz}
=&\,
-\frac{D^{3}}{12\pi^{2}v_{F}^{4}}
\Psi(\mu/D)\sum_{l=-\infty}^{\infty}j^{l}j^{l+1},\label{eq:res1}
\end{align}
where $D$ denotes the bandwidth of the TSS and
\begin{align}
\Psi(x)=2(1-x^{2})+\ln\left[\frac{1-x^{2}}{4}\right]
+x^{3}\ln\left[\frac{1+x}{1-x}\right].
\end{align}
This result is exemplary of what we will find later when including either hexagonal warping (Fig.~\ref{fig:cone}c) or the full momentum dependence of a finite-range potential (Fig.~\ref{fig:mucombi}), namely a symmetric function of $\mu$, which changes sign somewhere between $0$ and $D$. The perturbative result in~\eqref{eq:res1} has its zero at $\mu=0.483 D$, and the resulting anisotropy function is shown as the thick black curve in Fig.~\ref{fig:cone}c. We use parameters approximately relevant for Bi$_2$Se$_3$, and include only two finite harmonics, $2j^{1}=j^{0}= 10$ eV \AA$^{2}$, corresponding to a dimensionless coupling of $D j^{0}/v_{F}^{2}=0.44$.

Going beyond second order perturbation theory, we revert to a numerical diagonalization of the Hamiltonian $H=H_{el}+H_{sd}$ in momentum space, which is readily done when treating the impurity spin as a classical vector. We employ a quadratic momentum grid with spacing $dk=0.034 D/v_{F}$, with velocity $v_{F}\approx3000$ meV \AA,  and a TSS (half-)bandwidth of $D\approx400$ meV approximately relevant for Bi$_2$Se$_3$. From this grid we retain only momenta having $\varepsilon_{+\bk}<D$, and sum up the negative eigenvalues of the resulting matrix Hamiltonian, $H_{\bk'\bk}$, to find the ground state energies, $E_{0,i}$ for $i=x,y,z$, corresponding to different directions of a fixed impurity-spin \mbox{(unit-)vector}. From this, the longitudinal magnetic anisotropy is found as $\delta\Lambda_{zz}=E_{0,z}-E_{0,x}$. None of the parameter choices we have studied gave rise to any transverse anisotropy, i.e. $E_{0,x}=E_{0,y}$ to within the numerical precision, although the numercis do show hints at a very small contribution when including strong warping.

\begin{figure}[t]
\includegraphics[width=\columnwidth]{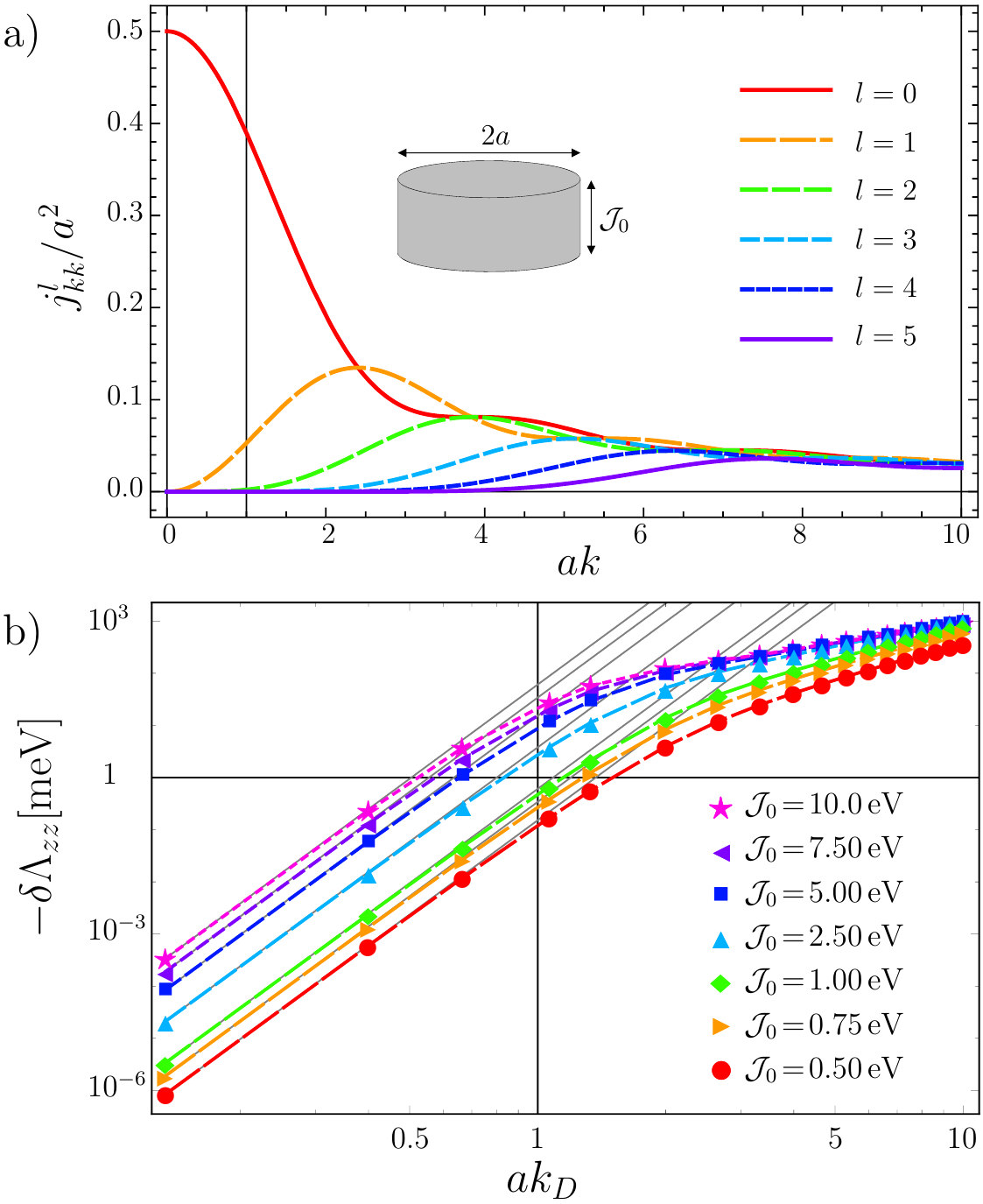}
\caption{(Color online) a) Plot of the maximum value, $j^{l}_{k k}$, of the 6 lowest harmonics of the isotropic step-function impurity potential, $\exch_{0}(r)$, with range $a$. b) Double-logarithmic plot of the corresponding longitudinal magnetic anisotropy energy at $\mu=0$ as a function of impurity range $a$ for a number of different amplitudes $\exch_{0}$. Curves in red to magenta (and smaller dashing) correspond to larger values of $\exch_{0}$. The range is measured in units of the smallest available wavelength of the TSS Dirac band, $k_{D}^{-1}=v_{F}/D=7.5$\AA. Thin gray lines show the power-law $0.77(W^{3}/v_{F}^{4})(a^{2}\exch_{0})^{2}(ak_{D})^{6}$, matching the curves very well for $ak_{D}\lesssim 1$.}
\label{fig:diskharm}
\end{figure}
Moving on to a concrete spatially isotropic, but finite-ranged impurity, we shall consider a step-function impurity exchange potential, $\exch_{0}(r)=\exch_{0}\theta(a-r)$, where $r=|{\bf r}|$ in two dimensions and $a$ sets the range of the potential. From this, one finds that $j^{l}_{k'k}\!=\exch_{0}a^{2}{\mathcal M}_{l}(ak,ak')$ with
\begin{align}
{\mathcal M}_{l}(x,y)=\frac{x J_{|l|-1}(x)J_{|l|}(y)-y J_{|l|-1}(y)J_{|l|}(x)}{y^{2}-x^{2}},
\end{align}
which attains its maximum at $k=k'$, with values $j^{l}_{kk}$ plotted in Fig.~\ref{fig:diskharm}a as a function of $ak$ for the six lowest harmonics. These may be inserted into the perturbative formula~\eqref{eq:dLamb} to determine $\delta\Lambda_{zz}$, but since this involves a 4-dimensional integral, it is in fact easier to carry out the numerical diagonalization, which is valid for any impurity strength $\exch_{0}$ and range $a$. To narrow down a relevant range of parameters, we first calculate the anisotropy energy at $\mu=0$ as a function of the potential range, $a$, for various potential strengths, $\exch_{0}$. The result is shown in Fig.~\ref{fig:diskharm}b to give $\delta\Lambda_{zz}(\mu=0)\sim -1$meV for $ak_{D}\sim 1$, meaning roughly that a sizeable effect ($\geq$1 meV) is to be expected for potentials of range comparable to or larger than the minimum wavelength, $k_{D}^{-1}=v_{F}/D\sim 7.5$ \AA$\,$  of available TSS electrons. The curves exhibit an initial power-law rise with $ak_{D}$, captured approximately by $-\delta\Lambda_{zz}(\mu=0)\approx0.77(W^{3}/v_{F}^{4})(a^{2}\exch_{0})^{2}(ak_{D})^{6}$ for $ak_{D}\lesssim 1$, and merge to roughly a $(ak_{D})^{1.3}$ dependence for $ak_{D}\gtrsim 1$. The exact dependence of the anisotropy on impurity range and strength relies on the specific impurity profile, $\exch_{0}(r)$, and for a Gaussian profile we have also confirmed the non-monotonous behavior found in Ref.~\cite{Nunez2012}, with a maximum anisotropy achieved for $ak_{D}\sim 1$.

We now return to the full dependence on chemical potential, which is shown in Fig.~\ref{fig:mucombi} for $\exch_{0}=1$ eV and $\exch_{0}=10$ eV and for varying impurity ranges. All the curves are seen to cross zero at the same value of $\mu$, and we note that for $\exch_{0}=1$ eV, but not for $\exch_{0}=10$ eV, the six curves collapse reasonably well onto one and the same curve, when scaling them with the power-law fit found above for $\mu=0$. For $\exch_{0}=10$ eV, the impurity potential shifts states outside of the unperturbed TSS band, corresponding to the shaded ranges with $|\mu|>D$, at which point it makes no sense to study the isolated TSS band, without the TI bulk bands from which they emanate. Including warping of the TSS band generally makes $\delta\Lambda_{zz}$ smaller and lowers the values of $|\mu|$ at which it changes sign, just as illustrated for the simpler example in Fig.~\ref{fig:cone}c, implying a more significant effect in Bi$_2$Se$_3$, which has weaker warping than Bi$_2$Te$_3$~\cite{Fu2009, Hasan2010}.

As we have demonstrated, a finite-range exchange coupling of a magnetic impurity with the TSS gives rise to a longitudinal magnetic anisotropy. The magnitude of this effect is found to be within the meV-regime, if the exchange coupling is of the order of 1 eV and its spatial range at least comparable to the smallest wavelength available in the TSS band, roughly 7.5 \AA, with the numbers used here for Bi$_2$Se$_3$ and omitting warping. In real systems, this {\it indirect} anisotropy will be added to the {\it intrinsic} anisotropy, $\delta\Lambda^{\rm intr}$, arising from crystal field splitting and atomic spin-orbit coupling and expected to be of the order of a few meV~\cite{Li2012, Eelbo2014}. Nevertheless, as we have demonstrated, the indirect term has a strong dependence on chemical potential, and even changes sign somewhere between half, and complete filling of the TSS band. This signature makes the two contributions experimentally discernible, and moreover it opens a pathway for controlling the adatom spin orientation, and thereby possibly the gap in the TSS Dirac spectrum, by tuning the chemical potential, either by doping or by electrical gating~\cite{Schlenk2013, Scholz2012, Steinberg2010, Zhang2013, Yang2015}.

The perturbative argument given above remains valid within a quantum mechanical treatment of the impurity spin, but already the third order correction will exhibit logarithmic corrections signifying the onset of Kondo effect for $\mu\neq 0$ when reducing temperature to a characteristic Kondo temperature, $T_{K}$~\cite{Zitko2010,Mitchell2013,Isaev2015}. The higher harmonics of a finite-range impurity makes this a potentially interesting multi-channel high-spin problem, but for strong enough intrinsic anisotropy, $|\delta\Lambda^{\rm intr}|>T_{K}$, Kondo effect is prohibited and one merely expects a downward renormalization of $|\delta\Lambda^{\rm in}|$~\cite{Otte2008, Zitko2008, Delgado2014}. On the other hand, for $|\delta\Lambda^{\rm intr}|<T_{K}$, the indirect contribution studied here, arising from the exchange coupling itself, will be enhanced by incipient Kondo effect until it grows large enough to impede the Kondo effect by itself. Settling this intricate balance between Kondo screening and indirect anisotropy spin-splitting is left for future studies, as is the obviously interesting extension of the present analysis to include a microscopic model for the exchange coupling of specific magnetic adatoms on a specific TI surface.

\textit{Acknowledgements}. We thank P. Kotetes and M. Misiorny for useful discussions. The Center for Quantum Devices is funded by the Danish National Research Foundation.

\bibliographystyle{apsrev4-1}
%

\end{document}